\documentclass[tightenlines,aps]{revtex4}
\usepackage{graphicx}
\usepackage{float}
\begin{document}
\title{Non--linear Fractal Interpolating Functions}
\author{R.~Kobes}
\email{randy@theory.uwinnipeg.ca}
\altaffiliation{ Winnipeg Institute for
Theoretical Physics, Winnipeg, Manitoba, R3B 2E9 Canada}
\affiliation{Department of Physics, University of Winnipeg, Winnipeg, 
Manitoba, R3B 2E9 Canada}
\author{ H.~Letkeman}
\affiliation{Department of Physics, University of Winnipeg, Winnipeg, 
Manitoba, R3B 2E9 Canada}
\begin{abstract}
We consider two non--linear generalizations of fractal
interpolating functions generated from iterated function systems. 
The first corresponds to fitting data
using a $K^{\rm th}$--order polynomial, while
the second relates to the freedom of adding certain arbitrary
functions. An escape--time algorithm that can be
used for such systems to generate fractal images like those associated
with Julia or Mandelbrot sets is also described. 
\end{abstract}
\maketitle
\section{Introduction}
An Iterated Function System (IFS) can be used to construct
a fractal interpolating function for a given set of data
\cite{barn,ifs}. The simplest such system defines an IFS
\begin{equation}
\left(\begin{array}{c}t^\prime\\ x^\prime\end{array}\right) =
\left( \begin{array}{cc} a_n & 0\\ c_n & 0\end{array} \right)
\left(\begin{array}{c}t\\x\end{array}\right) +
\left(\begin{array}{c}e_n\\f_n\end{array}\right),
\end{equation}
with coefficients $a_n, c_n, e_n$, and $f_n$ determined
from discrete data points ($t_i, x_i$), $i=0,1,\ldots, N$.
Such an IFS interpolates the data
set in the sense that, under certain assumptions on the coefficients
\cite{barn}, the attractor of the IFS is a graph that passes
through the data points. In this particular case, the IFS
can be written as 
\begin{eqnarray}
t^\prime &=&
\frac{(t-t_0)}{(t_N-t_0)}\ t_n
+\frac{(t-t_N)}{(t_0-t_N)}\ t_{n-1}
\nonumber\\
x^\prime &=&
\frac{(t^\prime-t_{n-1})}{(t_n-t_{n-1})}\ x_n
+\frac{(t^\prime-t_n)}{(t_{n-1}-t_n)}\ x_{n-1}
\end{eqnarray}
which shows that a linear (in $t$)
interpolating function between the points ($t_{n-1}, x_{n-1}$)
and ($t_n, x_n$) is used. 
\par
Various generalizations of fractal interpolating functions
have been given, including those for higher dimensional
functions, the use of hidden variables, and extensions to
certain non--linear distortions \cite{barn1,mass,kocic,nl,groller}.
In this note we describe a generalization whereby the transformation
incorporates a $K^{\rm th}$--order polynomial interpolation between
adjacent points. We also discuss certain classes
of non--linear functions that can arise in such interpolating
functions, and show how such functions can, with the use
of a particular escape--time algorithm, be used to generate
certain fractal images.
\par
The paper is organized as follows. In Section \ref{lin}
we describe simple linear fractal interpolating functions,
and discuss how particular non--linear functions can
arise. Section \ref{quadratic} generalizes these considerations
to $K^{\rm th}$--order interpolating functions. Section \ref{escape}
describes a certain escape--time algorithm which may be used
for these systems to generate fractal images like those
associated with Mandelbrot or Julia sets. Section \ref{end}
contains some brief conclusions.
\section{Linear Interpolating Functions}
\label{lin}
We first describe how a standard linear fractal 
interpolating function is constructed. Suppose we
have data points ($t_i, x_i$), $i=0\ldots N$,
describing a function $x(t)$. Consider the IFS
\begin{equation}
W_n\left(\begin{array}{c}t\\x\end{array}\right) =
\left( \begin{array}{cc} a_n & 0\\ c_n & 0\end{array} \right)
\left(\begin{array}{c}t\\x\end{array}\right) +
\left(\begin{array}{c}e_n\\f_n\end{array}\right)
\label{linear}
\end{equation}
Imposing the conditions, for $n=1,2,\ldots,N$,
\begin{eqnarray}
W_n\left(\begin{array}{c}t_0\\x_0\end{array}\right) &=&
\left(\begin{array}{c}t_{n-1}\\x_{n-1}\end{array}\right)
\nonumber\\
W_n\left(\begin{array}{c}t_N\\x_N\end{array}\right) &=&
\left(\begin{array}{c}t_{n}\\x_{n}\end{array}\right)
\label{lincond}
\end{eqnarray}
leads to determination of the coefficients as
\begin{eqnarray}
a_n &=& \frac{t_n - t_{n-1}}{t_N-t_0} \nonumber\\
e_n &=& \frac{t_{n-1}t_N - t_{n}t_0}{t_N-t_0} \nonumber\\
c_n &=& \frac{x_n - x_{n-1}}{t_N-t_0} \nonumber\\
f_n &=& \frac{x_{n-1}t_N - x_nt_{0}}{t_N-t_0}
\end{eqnarray}
The transformation can then be written as
\begin{eqnarray}
W_n(t) \equiv t^\prime &=&
\frac{(t-t_0)}{(t_N-t_0)}\ t_n
+\frac{(t-t_N)}{(t_0-t_N)}\ t_{n-1}
\nonumber\\
W_n(x) \equiv x^\prime &=&
\frac{(t^\prime-t_{n-1})}{(t_n-t_{n-1})}\ x_n
+\frac{(t^\prime-t_n)}{(t_{n-1}-t_n)}\ x_{n-1}
\end{eqnarray}
Thus, $W_n(x)\equiv x^\prime$ is determined by a linear
(in $t$) interpolating function
constructed between the points ($t_{n-1}, x_{n-1}$)
and ($t_n, x_n$).
\par
A generalization of this type of fractal interpolating
function can be found by considering an IFS of the form
\begin{eqnarray}
W_n(t) &=& a_n t + e_n\nonumber\\
W_n(x) &=& c_nt +f_n +g_n(x)
\label{ling}
\end{eqnarray}
where $g_n(x)$ is, at this stage, an arbitrary function. Imposing the 
conditions (\ref{lincond}) leads to determination of the
coefficients as
\begin{eqnarray}
a_n &=& \frac{t_n - t_{n-1}}{t_N-t_0} \nonumber\\
e_n &=& \frac{t_{n-1}t_N - t_{n}t_0}{t_N-t_0} \nonumber\\
c_n &=& \frac{x_n - x_{n-1}}{t_N-t_0} -
\frac{g_n(x_N)-g_n(x_0)}{t_N-t_0}\nonumber\\
f_n &=& \frac{x_{n-1}t_N - x_nt_{0}}{t_N-t_0}-
\frac{g_n(x_0)t_N-g_n(x_N)t_0}{t_N-t_0}
\end{eqnarray}
The transformation can then be written as
\begin{eqnarray}
W_n(t) \equiv t^\prime &=&
\frac{(t-t_0)}{(t_N-t_0)}\ t_n
+\frac{(t-t_N)}{(t_0-t_N)}\ t_{n-1}
\nonumber\\
W_n(x) \equiv x^\prime &=&
\frac{(t^\prime-t_{n-1})}{(t_n-t_{n-1})}\ {\tilde x}_n 
+\frac{(t^\prime-t_n)}{(t_{n-1}-t_n)}\ {\tilde x}_{n-1}
\end{eqnarray}
where 
\begin{eqnarray}
{\tilde x}_n &=& x_n +g_n(x)-g_n(x_N)\nonumber\\
{\tilde x}_{n-1} &=& x_{n-1} +g_n(x)-g_n(x_0)
\end{eqnarray}
and the identity
\begin{equation}
\frac{(t^\prime-t_{n-1})}{(t_n-t_{n-1})}
+\frac{(t^\prime-t_n)}{(t_{n-1}-t_n)} = 1
\end{equation}
for arbitrary $t^\prime$ has been used.
\section{Quadratic and Higher Order Interpolating Functions}
\label{quadratic}
The interpolating function of the last section used
a linear (in $t$) approximation between adjacent points.
In this section we indicate how a quadratic approximation
may be constructed; the generalization to an arbitrary 
$K^{\rm th}$--order polynomial approximation will be
straightforward.
Let us consider a transformation of the form
\begin{eqnarray}
W_n(t) &=& a_n t + e_n\nonumber\\
W_n(x) &=& c_nt +d_nt^2 +f_n
\end{eqnarray}
and impose the conditions, for $n=2,3,\ldots,N$,
\begin{eqnarray}
W_n\left(\begin{array}{c}t_0\\x_0\end{array}\right) &=&
\left(\begin{array}{c}t_{n-2}\\x_{n-2}\end{array}\right)
\nonumber\\
W_n\left(\begin{array}{c}t_m\\x_m\end{array}\right) &=&
\left(\begin{array}{c}t_{n-1}\\x_{n-1}\end{array}\right)
\nonumber\\
W_n\left(\begin{array}{c}t_N\\x_N\end{array}\right) &=&
\left(\begin{array}{c}t_{n}\\x_{n}\end{array}\right)
\label{quadcond}
\end{eqnarray}
The point $t_m$ is determined as
\begin{equation}
t_m = \frac{(t_{n-1}-t_{n-2})}{(t_n-t_{n-2})}\ t_N
+\frac{(t_{n-1}- t_{n})}{(t_{n-2}-t_{n})}\ t_0
\label{tm}
\end{equation}
with corresponding point $x_m$. The coefficients of the IFS are
determined as
\begin{eqnarray}
a_n &=& \frac{t_n-t_{n-2}}{t_N-t_0}\nonumber\\
e_n &=& \frac{t_Nt_{n-2}-t_0t_n}{t_N-t_0}\nonumber\\
c_n &=& \frac{x_n(t_0^2-t_m^2) +x_{n-1}(t_N^2-t_0^2)+x_{n-2}(t_m^2-t_N^2)}
{ (t_N-t_0)(t_N-t_m)(t_m-t_0)} \nonumber\\
d_n &=& \frac{x_n(t_m-t_0) +x_{n-1}(t_0-t_N)+x_{n-2}(t_N-t_m)}
{ (t_N-t_0)(t_N-t_m)(t_m-t_0)} \nonumber\\
f_n &=& \frac{x_nt_mt_0(t_m-t_0) +x_{n-1}t_Nt_0(t_0-t_N)
+x_{n-2}t_Nt_m(t_N-t_m)}
{ (t_N-t_0)(t_N-t_m)(t_m-t_0)}
\end{eqnarray}
With this, the transformation can be written as
\begin{eqnarray}
W_n(t) \equiv t^\prime &=&
\frac{(t-t_0)}{(t_N-t_0)}\ t_n
+\frac{(t-t_N)}{(t_0-t_N)}\ t_{n-2}
\nonumber\\
W_n(x) \equiv x^\prime &=&
\frac{(t^\prime-t_{n-1})(t^\prime-t_{n-2})}
{(t_n-t_{n-1})(t_n-t_{n-2})}\ x_n
+\frac{(t^\prime-t_{n})(t^\prime-t_{n-2})}
{(t_{n-1}-t_{n})(t_{n-1}-t_{n-2})}\ x_{n-1}\nonumber\\
&+&\frac{(t^\prime-t_{n-1})(t^\prime-t_{n})}
{(t_{n-2}-t_{n-1})(t_{n-2}-t_{n})}\ x_{n-2}
\end{eqnarray}
which thus uses a quadratic (in $t^\prime$) interpolating
function between the points ($t_n, x_n$), ($t_{n-1}, x_{n-1}$),
and ($t_{n-2}, x_{n-2}$).
\par
As in the previous section, including an arbitrary function
$g_n(x)$ in the IFS transformation via
\begin{eqnarray}
W_n(t) &=& a_n t + e_n\nonumber\\
W_n(x) &=& c_nt +d_nt^2 +f_n + g_n(x)
\label{quadg}
\end{eqnarray}
is straightforward. The conditions (\ref{quadcond}) leads
to determination of the point $t_m$ of 
Eq.~(\ref{tm}) as before, together with the accompanying
point $x_m$. The transformation itself can be written as
\begin{eqnarray}
W_n(t) \equiv t^\prime &=&
\frac{(t-t_0)}{(t_N-t_0)}\ t_n
+\frac{(t-t_N)}{(t_0-t_N)}\ t_{n-2}
\nonumber\\
W_n(x) \equiv x^\prime &=&
\frac{(t^\prime-t_{n-1})(t^\prime-t_{n-2})}
{(t_n-t_{n-1})(t_n-t_{n-2})}\ {\tilde x}_n
+\frac{(t^\prime-t_{n})(t^\prime-t_{n-2})}
{(t_{n-1}-t_{n})(t_{n-1}-t_{n-2})}\ {\tilde x}_{n-1}\nonumber\\
&+&\frac{(t^\prime-t_{n-1})(t^\prime-t_{n})}
{(t_{n-2}-t_{n-1})(t_{n-2}-t_{n})}\ {\tilde x}_{n-2}
\end{eqnarray}
where
\begin{eqnarray}
{\tilde x}_n &=& x_n +g_n(x)-g_n(x_N)\nonumber\\
{\tilde x}_{n-1} &=& x_{n-1} +g_n(x)-g_n(x_m)\nonumber\\
{\tilde x}_{n-2} &=& x_{n-2} +g_n(x)-g_n(x_0)
\end{eqnarray}
and the identity
\begin{equation}
\frac{(t^\prime-t_{n-1})(t^\prime-t_{n-2})}
{(t_n-t_{n-1})(t_n-t_{n-2})} +
\frac{(t^\prime-t_{n})(t^\prime-t_{n-2})}
{(t_{n-1}-t_{n})(t_{n-1}-t_{n-2})} +
\frac{(t^\prime-t_{n-1})(t^\prime-t_{n})}
{(t_{n-2}-t_{n-1})(t_{n-2}-t_{n})} = 1
\end{equation}
for arbitrary $t^\prime$ has been used.
\par
From these considerations, the pattern to
constructing a $K^{\rm th}$--order
fractal interpolating function is apparent.
Start with a transformation of the form
\begin{eqnarray}
W_n(t) &=& a_n t + e_n
\nonumber\\
W_n(x) &=& B_n^{(0)} + B_n^{(1)}t + B_n^{(2)}t^2 +\ldots + B_n^{(K)}t^K
\end{eqnarray}
and impose the conditions, for $n=K,K+1,\ldots,N$,
\begin{eqnarray}
W_n\left(\begin{array}{c}t_0\\x_0\end{array}\right) &=&
\left(\begin{array}{c}t_{n-K}\\x_{n-K}\end{array}\right)
\nonumber\\
W_n\left(\begin{array}{c}t_{m1}\\x_{m1}\end{array}\right) &=&
\left(\begin{array}{c}t_{n-K+1}\\x_{n-K+1}\end{array}\right)
\nonumber\\
W_n\left(\begin{array}{c}t_{m2}\\x_{m2}\end{array}\right) &=&
\left(\begin{array}{c}t_{n-K+2}\\x_{n-K+2}\end{array}\right)
\nonumber\\
&\vdots&\nonumber\\
W_n\left(\begin{array}{c}t_N\\x_N\end{array}\right) &=&
\left(\begin{array}{c}t_{n}\\x_{n}\end{array}\right)
\label{arb}
\end{eqnarray}
The $K-1$ intermediate points $t_{mj}$, with $j=1,2,\ldots, K-1$,
are determined as
\begin{equation}
t_{mj} = \frac{(t_{n-K+j}-t_{n-K})}{(t_n-t_{n-K})}\ t_N
+\frac{(t_{n-K+j}- t_{n})}{(t_{n-K}-t_{n})}\ t_0
\end{equation}
along with the corresponding $x_{mj}$ points.
The resulting transformation 
will be of the form given by Lagrange's formula for a
$K^{\rm th}$--order polynomial interpolating function 
constructed from $K+1$ points:
\begin{eqnarray}
W_n(t) \equiv t^\prime &=&
\frac{(t-t_0)}{(t_N-t_0)}\ t_n
+\frac{(t-t_N)}{(t_0-t_N)}\ t_{n-K}
\nonumber\\
W_n(x) \equiv x^\prime &=&
\frac{(t^\prime-t_{n-1})(t^\prime-t_{n-2})\cdots(t^\prime-t_{n-K})}
{(t_n-t_{n-1})(t_n-t_{n-2})\cdots(t_n-t_{n-K})}\ x_n  +\nonumber\\
&+&\frac{(t^\prime-t_{n})(t^\prime-t_{n-2})\cdots(t^\prime-t_{n-K})}
{(t_{n-1}-t_{n})(t_{n-1}-t_{n-2})\cdots(t_{n-1}-t_{n-K})}\ x_{n-1}
+\ldots+\nonumber\\
&+&\frac{(t^\prime-t_{n})(t^\prime-t_{n-1})\cdots(t^\prime-t_{n-K+1})}
{(t_{n-K}-t_{n})(t_{n-K}-t_{n-1})\cdots(t_{n-K}-t_{n-K+1})}\ x_{n-K}
\end{eqnarray}
The inclusion of an arbitrary function $g_n(x)$ in the
transformation $W_n(x)$ of Eq.~(\ref{arb}), as was done
for the linear and quadratic transformations of Eqs.~(\ref{ling}) 
and (\ref{quadg}) respectively, is straightforward. As might
be expected, the use of these higher--order interpolating
functions can increase the accuracy of the interpolation
significantly, at least
for smooth functions -- some informal tests on known
functions suggest an
improvement of almost an order of magnitude in general
in using a quadratic interpolating function over
a linear one. Of course, as for polynomial interpolation, 
there is a limit to the net gain in employing a
higher--order interpolating function.
\section{Escape--Time Algorithm}
\label{escape}
Assuming that the corresponding IFS transformation is contractive,
so that the distance $d(t, x)$ between any two points in the
range of interest satisfies
\begin{equation}
  d(W_n(t), W_n(x)) \le s_n\ d(t, x), 
\end{equation}
where $0 < s_n \le 1$ is the contractivity factor, 
graphs of the functions represented by fractal interpolating
functions can be made by applying the standard random iteration
algorithm to the IFS:
\begin{itemize}
\item {\tt initialize} ($t, x$) {\tt to a point in the interval of interest}
\item {\tt for a set number of iterations}
  \begin{itemize}
  \item {\tt randomly select a transformation} $W_n(t, x)$
  \item {\tt plot} ($t^\prime, x^\prime $) $ = W_n(t, x)$
  \item {\tt set} ($t, x$) $ = $ ($t^\prime, x^\prime $)
  \end{itemize}
 \item {\tt end for}
\end{itemize}
Alternatively, one can relate an IFS $W_n(t, x)$ to a shift 
dynamical system $f(t, x)$, and on this system perform an
escape time algorithm to generate an image \cite{barn}.
In this section we describe an algorithm for generating 
fractal images like those for Julia or Mandelbrot
sets from IFS interpolating functions.
\par
Suppose we have an IFS transformation $W_n(t, x)$, generated by some data
points ($x_i, y_i$), $i=0,1,\ldots, N$, which includes a
non--linear function $g_n(x)$, as was done
for the linear and quadratic transformations of Eqs.~(\ref{ling}) 
and (\ref{quadg}) respectively. We now continue the real variable
$x$ of this transformation to complex values: $x\to z = $ ($z_R, z_I$), 
so that the transformation $W_n(t, z)$ is defined on the complex plane.
We can then, in analogy with the algorithm used
for Julia sets, define the following escape--time algorithm
to generate a fractal pattern:
\begin{itemize}
\item {\tt for each pixel in a region of interest}
  \begin{itemize}
  \item {\tt initialize} $t$
   \item {\tt initialize} $z = $ ($z_R, z_I$) {\tt to the pixel coordinates}
  \item {\tt for} $n=0,1,\ldots, N$
    \begin{itemize}
      \item  {\tt calculate} ($t^\prime, z^\prime $) $ = W_n(t, z)$
      \item {\tt break if} $\sqrt{z_R^{\prime\,2} + z_I^{\prime\,2}}$ 
        {\tt exceeds a maximum}
      \item  {\tt set} ($t, z$) $ = $ ($t^\prime, z^\prime $)
    \end{itemize}
  \item {\tt end for}
  \item {\tt plot the pixel}
  \end{itemize}
\item {\tt end for}
\end{itemize}
where the pixel is plotted using a coloring algorithm
based upon, amongst perhaps other factors, the number of iterations
attained when the break condition was met \cite{color}. 
\par
The preceding can be interpreted as follows.
A general $K^{\rm th}$--order fractal interpolating IFS
\begin{eqnarray}
W_n(t) &=& a_n t + e_n
\nonumber\\
W_n(x) &=& B_n^{(0)} + B_n^{(1)}t + B_n^{(2)}t^2 +\ldots + B_n^{(K)}t^K
+ g_n(x),
\end{eqnarray}
with the coefficients $a_n, e_n, B_n^{(0)}, B_n^{(1)}, \ldots, B_n^{(K)}$
determined from the data ($t_i, x_i$), $i=0,1,\ldots, N$, can be
viewed, with the continuation $x\to z$, as defining a complex map
\begin{eqnarray}
t_{n+1} &=& a_n t_n + e_n
\nonumber\\
z_{n+1} &=& B_n^{(0)} + B_n^{(1)}t_n + B_n^{(2)}t_n^2 +\ldots + B_n^{(K)}t_n^K
+ g_n(z_n)
\end{eqnarray}
for $n=0,1,\ldots, N$. The escape--time algorithm described
above is then just the standard one used for
Julia sets of complex maps.
\par
The arbitrariness of the function $g_n(x)$ and the data set
($t_i, x_i$) used to fix the IFS leads to a wide variety of possible
fractal images generated in this way.
An interesting class of functions $g_n(x)$ to consider in this
context are those for which, when
continued to the complex plane $x\to z = $ ($z_R, z_I$), lead to
a map having a fixed point $z_I^* = 0$:
\begin{equation}
  z_I^* = 0 = {\rm Im} g_n(z_R, z_I^*)
\label{bo}
\end{equation}
In such a case one could augment the usual condition of the
escape--time algorithm to cease iteration:
$ \sqrt{z_R^2 + z_I^2} > \Lambda$, 
where $\Lambda$ is some suitably large value, to also cease
iteration when  $ |z_I| < \epsilon$,
where $\epsilon$ is a suitably small value. The coloring algorithm
used to plot a pixel, which depends on the number of iterations
attained when this break--out condition was met (if at all), will
then lead to structure in the region where the break--out
condition on the magnitude of $z$ is not met.
\par
\begin{figure}[H]
\begin{center}
\includegraphics[height=250pt]{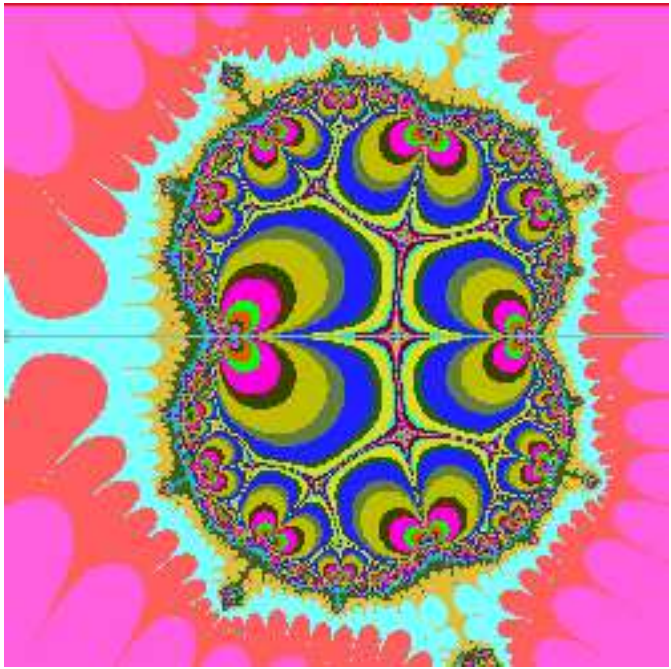}
\end{center}
\caption{Fractal image from a $t$--linear interpolating function}
\label{ifs1}
\end{figure}
\begin{figure}[H]
\begin{center}
\includegraphics[height=250pt]{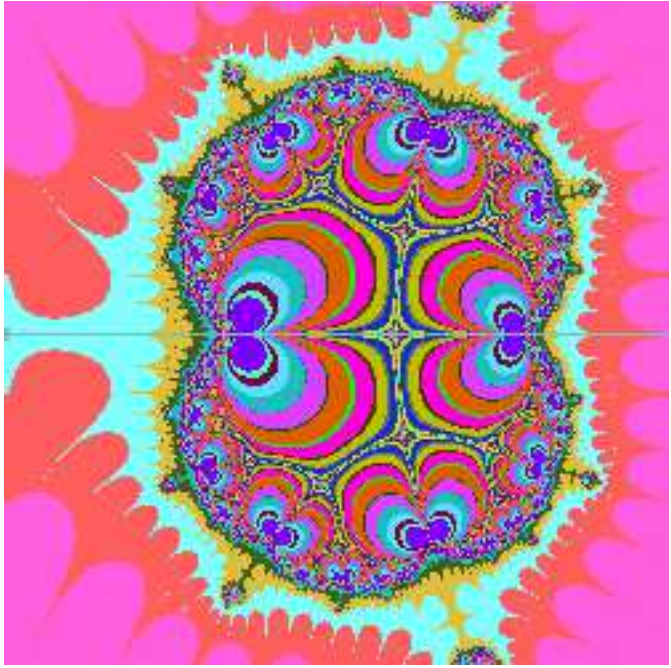}
\end{center}
\caption{Fractal image from a $t$--quadratic interpolating function}
\label{ifs2}
\end{figure}

We give two examples of fractal images generated this way
for the choice $g_n(x) =  -0.4x+0.5x^2+0.2x^3$, with the
data generated from the logistic map $x_{n+1} = 3.4x_n(1-x_n)$,
with $n=0,1,\ldots 60$. The first one, appearing in Fig.~\ref{ifs1},
corresponds to the generalization (\ref{ling}) of a 
linear (in $t$) fractal interpolating function, while the second image of
Fig.~\ref{ifs2} corresponds to the generalization (\ref{quadg}) 
of a quadratic (in $t$) interpolating function. A coloring algorithm
that simply mapped a color to the number of iterations attained
when the break--out condition became satisfied was used in
both cases. 
\par
\par
These figures illustrate, in the interior of the fractal object, 
the richer structure arising from the quadratic
over the linear interpolation function. 
In this region the break--out condition
$|z_I| < \epsilon$ is satisfied, which numerically for
$\epsilon \sim 10^{-5}$ is attained after a relatively small
number (10--30) of iterations.
\section{Conclusions}
\label{end}
We have considered two non--linear generalizations of
fractal interpolating functions constructed from iterated
function systems. One -- using a $K^{\rm th}$--order
interpolating polynomial -- can potentially improve the
accuracy of fractal interpolating functions. The other
generalization -- the use of certain arbitrary functions in
the IFS -- can, together with an appropriate escape--time
algorithm, generate fractal images. This last point
is of interest as, first of all, there is a rich variety of
such images possible due to the arbitrariness of the functions
used, and secondly, it shows how fractal images as normally
associated with Julia or Mandelbrot sets can also be associated
with discrete data sets.
\acknowledgments
This work was supported by the Natural Sciences and Engineering
Research Council of Canada.

\end{document}